\documentclass[11pt]{article}   \usepackage[round]{natbib}
\usepackage{amssymb} \usepackage{color}
\def\aj{AJ}% % Astronomical Journal  \def\actaa{Acta Astron.}% % Acta
\def\araa{ARA\&A}% % Annual Review of Astron and
\def\apj{ApJ}% % Astrophysical Journal  
\def\apjl{ApJ}% %
\def\apjs{ApJS}% % Astrophysical
\def\aap{A\&A}% % Astronomy and Astrophysics 
\def\mnras{MNRAS}% % Monthly Notices of
\def\pasp{PASP}% % Publications of the ASP  
\def\nat{Nature}% % Nature 
 
\def\aplett{ApL}

\newcommand{\trlx}{\mbox{${t_{\rm relax}}$}}
 \newcommand{\kms}{\mbox{${\rm
km~s}^{-1}$}} \newcommand{\msun}{\mbox{${\rm M}_\odot$}}

\def\apgt{\ {\raise-.5ex\hbox{$\buildrel>\over\sim$}}\ }  \def\aplt{\
{\raise-.5ex\hbox{$\buildrel<\over\sim$}}\ }  \def\lt{\
{\raise-.5ex\hbox{$\buildrel>$}}\ }  \def\gt{\
{\raise-.5ex\hbox{$\buildrel<$}}\ }

\def\Msun{\ensuremath{{\rm M}_{\odot}}}  \def\SgrA{Sgr\,A$^\star$}
\def\kms{\ensuremath{{\rm km}\,{\rm s}^{-1}}}
\usepackage{geometry} % to change the page dimensions
\begin{document}

\title{The MODEST questions: challenges and future directions in
stellar cluster research}

\author{Melvyn B. Davies$^1$, Pau Amaro-Seoane$^2$,  Cees Bassa$^3$,
Jim Dale$^4$, \\ Francesca De Angeli$^5$, Marc Freitag$^5$, Pavel
Kroupa$^6$, Dougal Mackey$^5$, \\ M. Coleman Miller$^7$, Simon
Portegies Zwart$^8$}

\maketitle

%\begin{center}
{\noindent $^1$Lund Observatory, Box 43, SE--221 00 Lund, Sweden; \\
\noindent $^2$Max-Planck Institut f\"ur Gravitationsphysik
(Albert-Einstein-Institut), Am M\"uhlenber 1, D--14476 Potsdam,
Germany; \\
\noindent $^3$Astronomical Institute, University of Utrecht, The
Netherlands; \\
\noindent $^4$Dept Physics \& Astronomy, University of Leicester,
Leicester, LE1 7RH, UK; \\
\noindent $^5$Institute of Astronomy, Madingley Road, Cambridge CB3
0HA, UK; \\
\noindent $^6$Argelander Institut f\"ur Astronomie (Sternwarte),
 Universit\"at Bonn, D-53121 Bonn, Germany; \\
\noindent $^7$Department of Astronomy, University of Maryland at
College Park, College Park, MD 20742-2421, USA; and Laboratory for
Gravtitational Astrophysics, NASA Goddard Space Flight Center,
Greenbelt, MD 20771, USA\\
\noindent $^8$Astronomical Institute "Anton Pannekoek" and Section
Computational Science, University of Amsterdam, Amsterdam, The
Netherlands.}
%\end{center} 

\medskip
\centerline{\sl Accepted for publication in New Astronomy}

\begin{abstract}
We present a review of some of the current major challenges in stellar
cluster research, including young clusters, globular clusters, and
galactic nuclei. Topics considered include: primordial mass
segregation and runaway mergers, expulsion of gas from clusters, the
production of stellar exotica seen in some clusters (eg blue
stragglers and extreme horizontal--branch stars),  binary populations
within clusters, the black--hole population within stellar clusters,
the final parsec problem, stellar dynamics around a massive black
hole, and stellar collisions.  The {\em Modest Questions} posed here
are the outcome of discussions which took place at the Modest-6A
workshop held in Lund, Sweden, in December, 2005.  Modest-6A was
organised as part of the activities of  the  {\em Modest
Collaboration} (see www.manybody.org for further details).
\end{abstract}
 
\section{Introduction}

MODEST is an abbreviation for MOdeling DEnse STellar systems, and is
a collaboration between groups working throughout the world on stellar
cluster research, including both theoreticians and observers.  The
Modest-6A workshop was held in Lund, Sweden, in 2005, as part  of the
continuing activities of the Modest Collaboration. A particular task
of this workshop was to produce a list of challenges in stellar
cluster research -- {\em The Modest Questions} -- considering both
problems likely to be solved in the shorter term (around one year) and
those requiring more work (timescales of several years).  This paper
provides a review of what came out of the discussion at
Modest--6A. The topics have been grouped into three areas (young
clusters, globular clusters, and galactic nuclei) although there is
naturally some overlap between the three sections.

\section{Young Clusters}

 \subsection{Primordial mass segregation and runaway mergers}
 \indent Dynamical interactions between stars in young clusters leads
 to mass segregation, in which the heavier stars sink towards the
 cluster centres \citep{1998MNRAS.295..691B,2002MNRAS.337..597D,
 2004NewAR..48...47K,2006MNRAS.369.1392F}.  This occurs very rapidly,
 on a timescale of roughly $t_{\rm mseg} \approx t_{\rm relax} m_{\rm
 av}/m_{\rm heavy}$, where $t_{\rm relax}$ is the two-body relaxation
 time of the cluster and $m_{\rm av}, m_{\rm heavy}$ are the average
 stellar mass and the mass of the heavy stars; $t_{\rm mseg}$ may
 approach the crossing time for very young dense clusters.  However,
 clusters may \textit{form} in a mass-segregated state by the
 competitive accretion process outlined in \cite{1997MNRAS.285..201B}.
 Whether or not mass segregation is a primordial state of star
 clusters is a crucial question because it affects the timescale for
 core collapse and it also has some bearing on the origin of
 massive stars.  This in turn affects the process of mergers involving
 massive stars in cluster cores. A cluster which is intially
 mass-segregated is more likely to undergo a runaway merger process,
 since the massive stars are already situtated at the cluster centre.\\

In a star cluster where the two-body relaxation time is sufficiently
small ($\trlx \aplt 100$\,Myr) the most massive stars can reach the
cluster core by dynamical friction and drive the cluster to a state of
core collapse before they explode
\citep{2002ApJ...576..899P,2004ApJ...604..632G,2005astro.ph..3130F}.
During this phase the stellar number density in the cluster core
becomes so high ($n_c \apgt 10^8$\,stars/pc$^3$) that stars may
experience direct physical collisions. This may lead to a 
runaway collision process 
in which one star repeatedly collides with other stars
\citep{1999A&A...348..117P}. The growth rate of this object may exceed
$10^{-3}$\,\msun/yr, and can therefore exceed the stellar mass loss
rate, which for the most massive stars is of the same order
\citep{2001A&A...369..574V}. The mass of the single massive object may
grow to about 1000--3000\,\msun\, although the subsequent evolution of
this object is unclear; it may for example collapse to form a black
hole \citep{2004Natur.428..724P}.  Observations may tell us whether or
not such intermediate mass black holes form in dense stellar
clusters. Good candidates so far are the young and dense star cluser
MGG-11 in the starburst galaxy M82 \citep{2003ApJ...596..240M}, and
the stellar conglomerate IRS13E very near the Galactic center
\citep{2004A&A...423..155M}. The latter object is of particular
interest as its black hole may be of the order of 1500\,\msun\, to
15000\,\msun\, \citep{2005ApJ...625L.111S}. The subject of intermediate-mass
black holes within globular clusters will be discussed in section 3.7.

\medskip

\indent For the MODEST questions and tasks on this subject, we propose:
 \begin{itemize}
 \item In the next year: perform N-body simulations of clusters with
 and without primordial mass segregation to determine its effect on
 the evolution of the clusters, particularly in regard to core
 collapse and runaway mergers.

% Recommend the following question be deleted, since it is essentially the same %as the second and third questions in section 3.7 

%\item In the next three years: Is there observational evidence for the
%      existence of IMBHs in star clusters?
 \item In the next three to ten years: Do any mass--segregated systems
 exist which are too young to have segregated dynamically
 \citep{2004A&A...416..137G}?
\item In the next ten years: How do winds and instabilities affect the
evolution of massive stars and what implications does this have for
the behaviour of very massive merger products?
\end{itemize}

%Original questions for this section are below:

%\indent For the MODEST questions on this subject, we propose:
% \begin{itemize}
% \item In the next year: perform N-body simulations of clusters with
% and without primordial mass segregation to determine its effect on
% the evolution of the clusters, particularly in regard to core
% collapse and runaway mergers.
%\item In the next three years: look for observational evidence for the
%      existence of IMBHs in star clusters.
% \item In the next three to ten years: conduct an observational search
% for segregated systems which are too young to have segregated
% dynamically \citep{2004A&A...416..137G}.
%\item In the next ten years: understand, from a theoretical and
%      observation point of view the properties of very massive stars,
%      in particular their winds and instabilities, and use this to
%      constrain the likely behaviour of very massive merger products.
%\end{itemize}

 \subsection{Gas expulsion from clusters}
 \indent Embedded clusters represent a crucial but poorly-understood
 phase in the process by which a giant molecular cloud is converted to
 a population of stars
 \citep{2005MNRAS.358..291D,2005MNRAS.359..809C}.  The observation
 that most embedded clusters do not survive to become open or globular
 clusters \citep{2003ARA&A..41...57L} implies either that most
 embedded clusters become unbound during the star-formation process,
 or that the giant molecular clouds from which they form were never
 bound in the first place \citep{2000ApJ...530..277E}.

 There are essentially three types of embedded systems
 \citep{2002MNRAS.336.1188K}: {\it Type~I} contain from a few to~1000
 stars but no O~stars since they are too rare to be sampled from the
 IMF. {\it Type~II} have between about $10^3$ and $10^5$ stars and
 contain between a few to about one hundred O~stars. {\it Type~III}
 clusters are massive with $\gtrsim 10^5$~stars.  Gas-expulsion may
 take a few crossing times for type~I clusters because the cumulative
 feedback energy from low-mass stars drives the gas out. For type~II
 clusters gas expulsion may be explosive because the O~stars provide
 sufficient feedback energy to blow out the gas on a crossing 
timescale or shorter.  In the very massive type ~III systems, feedback
 from O-stars is likely to be insufficient to remove gas until the
 detonation of the first supernovae because the gas density in these
 systems is large enough to quench the effects of photoionisation and
 winds. Such massive systems can therefore achieve core collapse while
 still containing substantial quantities of gas and therefore
 modelling them poses a particularly difficult problem  as one must
 allow for  both the dynamical N-body  and hydrodynamical evolution of
 the system.

The efficiency with which feedback expels gas determines the
 star-formation efficiency and also the likelihood of the cluster
 becoming unbound: a cluster unbinds more readily for low star
 formation efficiencies and/or for gas expulsion timescales shorter
 than a crossing time \citep{1984ApJ...285..141L,1997MNRAS.284..785G}.
 It is therefore important to model this process correctly. Gas
 expulsion also converts a hydrodynamical problem into an N-body
 problem, and thus determines when N-body calculations may start
 \citep{2001MNRAS.323..988G,2001MNRAS.321..699K}. At present, stellar
 feedback is poorly understood because it is not possible to treat
 directed outflows self-consistently, nor is it possible to handle the
 full radiative transport problem in three dimensions.  \\

\medskip

\indent We propose the following MODEST questions and tasks:
 \begin{itemize}
\item In the next one to three years: creation of a hybrid N-body SPH
      code, using high-precision N-body codes such as NBODY6 or {\tt
      starlab} as a starting point, since most existing hybrid codes
      are SPH codes which have had an N-body component grafted on.

\item In the next ten years: What is the efficiency with which O-stars
expel gas from typical clusters and hence what is the typical
efficiency of star-formation?
\item In the next ten years: study the formation of a
$10^{6}M_{\odot}$ cluster from the collapse of its natal molecular
cloud until the stage in cluster evolution where mass loss becomes
dominated by stellar evolution
\end{itemize}

%Original questions for this section are below

%\indent We propose the following MODEST questions:
% \begin{itemize}
%\item In the next one to three years: creation of a hybrid N-body SPH
%code, using an advanced N-body code such as NBODY6 as a starting
%point, since most existing hybrid codes are SPH codes which have had
%an N-body component grafted on.
%\item In the next ten years: estimate from hydrodynamical simulations
%including feedback the efficiency with which O-stars expel gas from
%typical clusters and hence determine typical star-formation
%efficiencies.
%\item In the next ten years: study the formation of a
%$10^{6}M_{\odot}$ cluster from the collapse of its natal molecular
%cloud until the stage in cluster evolution where mass loss becomes
%dominated by stellar evolution
%\end{itemize}

 \subsection{Cluster complexes}
\indent It has been recognized for some time that studying the
formation of isolated stars does not paint a realistic picture of star
formation, since virtually all stars form in clusters and many
interact with each other during their formation. There is also a
considerable body of evidence that cluster formation is itself
clustered. Interacting galaxies have been known for some time to host
vast networks of very massive star clusters whose formation is
triggered by mergers or tidal interactions, e.g. the Antennae
\citep{1995AJ....109..960W} and M82
\citep{2001AJ....121..768D}. However, as pointed out by
\cite{1999A&A...345...59L} the formation of such cluster associations
is not confined to merging galaxies. Dwarf galaxies (e.g. NGC 1569 and
NGC 1705, \cite{1994ApJ...433...65O}) and undisturbed spiral galaxies
(e.g. NGC 253, \cite{1996AJ....112..534W}, M101,
\cite{1996AJ....112.1009B}) also exhibit very large star--forming
complexes. \cite{1983MNRAS.203...31E} identified regularly--spaced
strings of giant HII regions or HI clouds in the spiral arms of 22
galaxies, finding that individual concentrations of star--formation
had typical size and mass scales of $1$--$4$ kpc and
$10^{6}$--$10^{7}$ M$_{\odot}$ respectively, considerably larger than
the sizes and masses of typical molecular clouds or OB associations.\\
\indent Star--formation in the Milky Way itself also appears to occur
on scales much larger than traditionally
assumed. \cite{1979SvA....23..163S} identifies more than ten
indicators of star formation and lists 49 star--formation regions with
masses of up to $\sim10^{6}$ M$_{\odot}$. \cite{1978SvAL....4...66E}
used Galactic Cephied variables to identify 35 complexes of star
formation with sizes of $\sim600$ pc and \cite{1989SvA....33..596B}
used spatial and kinematic data on 69 open clusters to identify eleven
star complexes with sizes ranging from $\sim10^{2}$--$\sim10^{3}$
pc. \cite{1979SvAL....5...12E} suggests that the sizes of these
regions simply reflect the original sizes of the complexes of gas and
dust from which they formed. He then proposes that the spatial and
temporal distribution of star--formation within the complexes is a
result of self--propagating star--formation driven by feedback from
O--stars.\\ \indent The study of star cluster complexes is therefore
of great importance, since it will shed light on the importance of
triggering in star formation on a variety of lengthscales. The
evolution of such complexes is clearly an important factor in the
formation and evolution of individual star clusters
\citep{1998MNRAS.300..200K} and may also have a bearing on the origins
of Ultra-Compact Dwarf galaxies and `faint fuzzies'
\citep{2002MNRAS.330..642F,2002AJ....124.2006F,2005ApJ...630..879F}.\\

%Original section is below:

 %\indent It has been recognized for some time that studying the
 %formation of isolated stars does not paint a realistic picture %of
 %star formation, since virtually all stars form in clusters and %many
 %interact with each other during their formation. Recently, the
 %discovery of vast complexes of star formation in interacting
 %galaxies %has raised the question of whether cluster formation is
 %itself %clustered and, if so, what effect this has on the formation
 %and %evolution of star clusters (Kroupa 1998). The evolution of these
 %cluster complexes may have a bearing on the formation of
 %Ultra-Compact Dwarf galaxies and `faint fuzzies'
 %\citep{2002MNRAS.330..642F,2002AJ....124.2006F,2005ApJ...630..879F}.\\

\medskip

\indent We propose the following MODEST questions:
\begin{itemize}
\item In the next year: Is star \textit{cluster} formation itself
always clustered, or do some clusters form and evolve alone?
\item In the next three years: From an observational perspective, how
do cluster complexes evolve and what is the role of triggering (on any
scale) in their formation?
\item In the next ten years: What bearing, if any, does the evolution
of star cluster complexes have on the formation of the various
dwarf-galaxy populations and the large faint star clusters (faint
fuzzies)?
\end{itemize}

\section{Globular Clusters}

\subsection{The production and evolution of blue stragglers}

Blue stragglers are main-sequence stars which are more massive than
the current turn-off mass. They have been seen in  globular and open
clusters, and in the halo.  They are believed to form in two ways:
either through collisions between lower-mass main-sequence stars
(which will be relatively frequent in dense stellar clusters
\nocite{1976ApL....17...87H} (Hills \& Day, 1976)), 
or via mass transfer within binaries
\citep{pre00,pio04,dav04,san05}.

%Little work has been done yet to model mass transfer within binaries,
% mostly because of the difficulty in following the evolution of a
% hydrodynamic system (the mass transfer between binary components) for
% the require length of time required (expected to be
% of the order of $10^9$ yr).

The subsequent evolution of merger and mass-transfer products remains
an open question.  Hydrodynamic calculations suggest that the
collision remnants do not develop substantial convective regions
during their thermal relaxation and therefore are not mixed
significantly after the collision \citep{sil97b}.   If true, this
would have a significant impact on the subsequent evolution of the
merger product, as the core would receive only a small amount of
hydrogen and would, therefore, have a relatively short life. However,
another key ingredient is the angular momentum contained in the merger
product.  Even a small, non-zero, impact parameter can result in an
object with substantial angular momentum \citep{sil02}.  This poses a
problem for its subsequent evolution. The total angular momentum can
be up to ten times larger than that possessed by
low-mass pre-main-sequence stars. The
collision product is expected to have quite a large radius soon after
the merger and to then  gradually contract back towards the
main-sequence. However if no angular momentum were removed during the
process, the object would reach the breakup velocity long before it
reaches the main-sequence. Some mechanism must remove most
($\approx$99\%) of the angular momentum. In pre-main-sequence stars
angular momentum loss is driven by surface convection zones and magnetic
winds, which are not expected to be present in collision
products. Recently \citet{dem04} inferred from spectroscopic data the
existence of circumstellar discs around 6 stars in a sample of 50
objects located above the main-sequence in 4 globular clusters.  The
presence of magnetically locked discs had already been suggested as a
possible mechanism to lose angular momentum \citep{leo95}. The
observed discs might not be massive enough to explain such an angular
momentum sink, but they could be the leftovers of once larger discs.

Unfortunately very little observational data for BSS rotation rates in
globular clusters currently exists.  Only a few BSS in globular
clusters have measured rotation so far
\citep{2005ApJ...632..894D}. Comparison between model predictions and
observation can thus be made only on the basis of colour-magnitude
diagrams. For example, \citet{sil05}  claim that  disc-free models of
BSS are brighter and bluer than the objects observed so far and
therefore imply that some angular momentum loss mechanism must be at
work.

\medskip

\indent We propose the following MODEST tasks:
\begin{itemize}
\item In the next one to three years: spectroscopic observations of
BSS will shed some light on the origin of BSS and on their subsequent
evolution. Rotation rates, surface gravity and chemical abundances are
fundamental information for probing collision models. Collisional BSS
are expected to form at the centre of globular clusters. For this
reason observing them spectroscopically will be quite challenging and
will probably require instruments such as HST/STIS and ground-based
adaptive-optic systems.
\item In the next one to three years: improve models of rotating
collisional products, and develop more detailed modelling of mass
transfer in binaries.  Predicted differences in the observables can
help distinguishing collisional from primordial BSS.
\end{itemize}

\subsection{Understanding the observed properties of extreme 
horizontal-branch stars}

% Francesca : changed the following paragraph
%Extreme horizontal-branch (EHB) stars are considerably bluer than
%regular horizontal-branch stars, owing to their much lower
%envelope masses.  EHB stars have been observed in several
%globular clusters.
%One of the mechanisms believed to form EHB stars involves a large
%envelope mass loss enhanced by tidal interaction within a close
%binary. 
Extreme horizontal-branch (EHB) stars have been observed in several
globular clusters as a group of objects considerably bluer than
regular horizontal-branch stars. It is by now widely accepted that EHB
stars are He burning stars that during their evolution have suffered
heavy mass loss \citep{ibe70,fau72}, keeping only a thin envelope
(with mass of the order of 0.02 M$_\odot$). However the actual
formation  mechanism for such object is still unclear.  Mass loss
during the horizontal branch (HB) phase has been proposed
\citep{wil84, yon00}, as well as enhanced mass loss rates during the
previous red giant branch (RGB) phase \citep{sok01}, through several
mechanisms.  One of these involves binarity, i.e. tidal interactions
within a close binary could enhance the envelope mass loss
\citep{men76,heb02,han02}.  However, preliminary results from
spectroscopic data \citep{mon06} show a lack of binaries among EHB
stars in NGC~6752, in sharp contrast with the results of
\citet{pet02}, that concluded that the majority of EHB stars in this
cluster are in binary systems. Apparently the two catalogues sample
different regions in the cluster, the \citet{pet02} one being located
in more external regions. This could imply a different formation
mechanism for EHB in the central and in the outer regions as has been
observed for BSS in many clusters by looking at the luminosity
functions \citep{fer04}. Recent results on the very massive clusters,
$\omega$~Cen \citep{pio05} and NGC~2808 \citep{dan05}, indicate that
the EHB stars in these two clusters could be the product of a second
generation of stars formed by material enriched in He due to the
pollution by SN and/or intermediate mass AGB stars. This would also
explain the correlation found between the extension of the EHB and the
mass of the clusters \citep{rec06}.

\medskip

\indent We propose the following MODEST question and task:
\begin{itemize}
\item In the next one to three years: could dynamical interactions in
dense stellar clusters trigger the large envelope mass loss believed
to be at the orgin of EHBs?
\item In the next one to three years: understand the binary
distribution among EHB stars by observing other clusters.
\end{itemize}

\subsection{Multiple episodes of star formation in some globular clusters?}

Photometric studies of red giants in $\omega$~Cen have revealed
several discrete populations covering the metallicity range  $-1.5
\leq [$M$/$H$] \leq -0.5$ with, possibly, an age spread of up to $\sim
6$ Gyr \citep{sol05}. Additional observations have revealed the
presence of a double main sequence, with a population of stars lying
to the blue of the primary main sequence \citep{bed04}. Spectroscopic
follow-up measurements provided the surprising result that the blue
main sequence is $\sim 0.3$ dex more  {\it metal-rich} than the red
population \citep{pio05}, the most likely explanation being that the
blue main sequence represents a super He-rich population of cluster
members. This interpretation is interesting because $\omega$~Cen also
possesses an EHB, for which one viable explanation is a population of
He-rich stars.

A number of other globular clusters are also known to possess unusual
stellar populations. Recent HST observations of NGC~2808 have
demonstrated this cluster to also have a population of blue main
sequence objects, although the main sequence does not show the clear
bifurcation present in $\omega$~Cen \citep{dan05}.  Again, this
population has been interpreted as He-rich -- a scenario which may
also help explain NGC~2808's EHB.  In addition, two Galactic bulge
clusters (NGC~6388 and 6441) also possess EHBs despite their rather
high metallicities ($[$Fe$/$H$]\sim -0.5$). Helium enhancement  has
been invoked to explain the anomalous HBs in these two objects. M54,
which lies near the centre of the Sagittarius dSph galaxy, is
suspected to possess  a small metallicity spread \citep{dac95}, and
has recently been shown to possess an EHB \citep{ros04}. Finally, the
very metal-poor remote halo cluster NGC~2419  also has an EHB. The one
common property of these disparate clusters is that they are all among
the most massive objects in the Galactic globular cluster system.

Both metallicity spreads and He-enhanced populations imply multiple
episodes of  star formation and self-enrichment in some globular
clusters, likely involving Type II supernovae and/or winds from
massive AGB stars. This picture is consistent with the observed
correlation with mass, as only the most massive clusters are likely to
be able to retain significant amounts of ejected gas. For example, the
scenario proposed by \citet{dan05} to explain the NGC~2808 main
sequence and EHB  proposes three distinct episodes of star formation
spread over several hundred Myr:  the initial burst at big-bang He
abundance ($Y\sim 0.24$), followed by a second  generation with $Y\sim
0.4$ born from the winds of massive ($\sim6-7 M_\odot$)  first
generation AGB stars, and later a third generation with $Y\sim
0.26-0.29$  born from the winds of less massive ($\sim3.5-4.5
M_\odot$) AGB stars. Similar  scenarios have been proposed for
$\omega$~Cen, although these often also include the possibility that
this cluster was formerly the nucleus of a now-defunct dwarf
galaxy. Numerous problems remain to be solved with self-enrichment
models,  involving, for example, the retention and mixing of ejected
gas within a cluster, the required first-generation IMFs (to get
enough AGB stars), and heavy-element  pollution from Type II supernove.

\medskip

We propose the following MODEST question and task:
\begin{itemize}
\item{In the next one to three years: can material enrichment take
place as a result of multiple mergers in very massive globular
clusters?}
\item{In the next three years: development of more comprehensive
models of self-enrichment (including hydrodynamic modelling
of accretion from interstellar gas within clusters) and multiple star-formation
episodes within globular clusters.}
% with the aim of self-consistently
%explaining the observed properties of objects such as $\omega$~Cen and
%NGC~2808. New and more detailed  observations, both photometric and
%spectroscopic, should follow  on a similar time-scale. These will be
%important for shaping and constraining the  self-enrichment models.}
\end{itemize}

\subsection{Observing stars escaping from globular clusters}
Measuring the properties of stars which are escaping from a particular
globular cluster has the potential to tell us much about the internal
processes in that cluster (eg, Gunn \& Griffin, 
1979\nocite{1979AJ.....84..752G}).  In
particular, it would be of interest to try and find stars which have
been ejected from the cluster with some significant velocity, as these
offer a means of probing strong interactions between cluster members,
such as three-body and four-body encounters (eg,
Meylan et al, 1991\nocite{1991ApJ...383..587M}).  
One possible way to locate such stars
would be by measuring proper motions in a nearby target (e.g. M4) near
the tidal radius or Lagrangian points. Once suitable candidates have
been located, radial velocities could also be obtained. Because proper
motion measurements require multi-epoch imaging over a significant
baseline, this is necessarily a problem to be tackled on a time-scale
of at least several years. Modelling can be utilized to predict the
number of expected detections, and will be vital in attempting to
constrain the processes which could produce the observed properties of
any strong candidate high-velocity escapers.  It is worth noting that
for several Galactic globulars (e.g., Palomar 5, 
Odenkirchen et al, 2003; Leon et al, 2000), \nocite{oden03}
\nocite{2000A&A...359..907L} large numbers of escaped stars have
been observed in the form  of tidal tails. These are members which
have drifted through the Lagrangian regions  with small relative
velocity, and are more useful for probing the cluster's orbit  about
the Galaxy as well as the Galactic mass distribution.

\medskip

We propose the following MODEST questions:
\begin{itemize}
\item{In the next three years: can we locate new tidal tails belonging
to any Galactic globular clusters? Searches utilizing deep wide-field
imaging are  presently in progress.} 
\item{In the next three years:
from a modelling aspect, can we
predict the number of expected detections of fast escapers from a
given globular cluster? Can we also constrain the processes which
produce them, and predict their observational properties (e.g., the
velocity distribution of fast escapers)?}
\item{In the next ten years: can we locate high proper motion stars
near the tidal boundaries of nearby globular clusters? If so, radial
velocities of any candidate stars need to be obtained. What can the
observed properties of such objects tell us about the internal
processes in their parent clusters?}
\end{itemize}

\subsection{Observational constraints on binary star 
populations in globular  clusters}

Relatively little is known about binary star populations in globular
clusters.   Photometrically, we can begin to deduce probable binary
candidates from the binary main sequence (eg, Rubenstein and Bailyn
1997)\nocite{1997ApJ...474..701R}.  This is a region of the CMD lying
above and to the red of the single-star main sequence. An unresolved
binary consisting of two main  sequence stars will have a combined
colour somewhere in between the colours of the two components, and a
magnitude brighter than that of the single-star main sequence at this
combined colour.  In principle, it is therefore possible to determine
cluster binary  fractions by observing the binary second sequence;
however in practice this process is complicated by photometric errors,
which mimic the main sequence spread due to binary stars, as well as
crowding and field star contamination.  Hence, to date binary
fractions have only been measured at low significance in Galactic
globular clusters via this method. Photometric variability surveys are
also sensitive to some types of cluster binaries -- in particular
those which have the correct inclination to be eclipsing objects, plus
those which are active in some manner, for example  contact binaries
(eg, Kaluzy et al. 1999\nocite{1999A&A...350..469K}).  However, the
relationship (if any exists) between active binaries and the global
population is not known, so it is difficult, if not impossible, to
infer properties of the normal binary population from the active one.

Cluster binaries can also be detected via spectroscopic observations
(eg, Pryor et al 1989 \nocite{1989ddse.work..175P}; Hut et al.  1992
\nocite{1992PASP..104..981H}).  The idea  is to obtain radial velocity
measurements at multiple epochs. Binary stars should show large
variations in the measured radial velocity due to the orbital motions
of the two components. Such measurements have the advantage that they
can provide information about the orbital period,  providing the
sampling is suffiently frequent. The disadvantage of this technique is
that it requires repeated time-consuming observations, and is rather
inefficient considering the binary fraction is expected to be of order
$10\%$ or less.

A third way of probing a cluster's binary star population is by means
of  X-ray observations.
%Paragraph from Cees inserted here:
With the launch of \emph{Chandra} and \emph{XMM--Newton}, many new
 X-ray sources have been detected in Galactic globular clusters. From
 all \emph{ROSAT} observations, 57 X-ray sources were discovered in as
 many globular clusters (Verbunt 2001\nocite{2001A&A...368..137V}),
 but in 47\,Tuc \emph{Chandra} already found over 300 sources
 (Grindlay et al. 2001, Heinke et
 al. 2005\nocite{2001Sci...292.2290G,2005ApJ...625..796H}). From
 \emph{Chandra} observations of 12 globular clusters, Pooley et
 al. (2003\nocite{2003ApJ...591L.131P}) found that the number of X-ray
 sources with an X-ray luminosity above
 $4\times10^{30}$\,erg\,s$^{-1}$ scales with the collision
 number. This scaling was interpreted as evidence that these X-ray
 sources, which are expected to be primarily cataclysmic variables,
 are formed through dynamical interactions.  From the X-ray colours
 and luminosities it is possible to identify and classify the sources
 containing an accreting neutron star (a low-mass X-ray binary), but
 for the large part of the X-ray sources optical identifications are
 necessary to discriminate cataclysmic variables from magnetically
 active binaries (RS\,CVn, BY\,Dra, W\,UMa systems). Sofar, this has
 been done for 4 globular clusters; NGC\,6752 (Pooley et
 al. 2002\nocite{2002ApJ...569..405P}), 47\,Tuc (Edmonds et
 al. 2003ab\nocite{2003ApJ...596.1177E,2003ApJ...596.1197E}),
 NGC\,6397 (Grindlay et al. 2001\nocite{2001ApJ...563L..53G}) and M4
 (Bassa et al. 2004\nocite{2004ApJ...609..755B}). These observations
 suggest that the majority of the X-ray sources with X-ray
 luminosities above a few times $10^{30}$\,erg\,s$^{-1}$ are
 cataclysmic variables, while the fainter sources are active
 binaries. From a comparison of these identifications in M4, 47\,Tuc
 and NGC\,6397, it is found that the number of such active binaries
 appears to scale with the (core) mass of the cluster, instead of the
 collision number (Bassa et al. 2004\nocite{2004ApJ...609..755B}). 
 Hence it seems likely that these systems evolved from primordial systems.

\medskip

We propose the following MODEST questions and tasks:
\begin{itemize}
\item{In the next year: can we make any robust estimates of binary
star fractions in globular clusters for which high-quality main
sequence photometry  already exists? This will require sophisticated
statistical treatment of observational  errors to disentangle the
binary second sequence. Does the spatial distribution of binaries look
like what is predicted from dynamical models of star clusters
(eg, Ivanova et al 2005)? \nocite{2005MNRAS.358..572I}}
\item{In the next three years: the establishment of a detailed
observing program of one nearby cluster (possibly M4), with the aim of
determining the binary fraction, along with the distribution of mass
ratios, periods, and separations for these objects. This will require
both photometry and multiple-epoch spectroscopic measurements. In
addition, can we start to establish a link between the active binary
population in this cluster, and the overall population? This would
allow us to start inferring the properties of binary populations in
other clusters based on already-existing observations of the active
systems.}
\item{In the next ten years: if the M4-type measurements can be
extended to several other clusters, we can start to build a global picture
of binary populations in globulars, including intra-cluster
variations. How do the characteristics of the binary populations vary
with cluster properties (e.g., mass, concentration, etc)? What can
this tell us about internal cluster dynamics?}
\end{itemize}

\subsection{Constraining initial conditions for globular cluster 
simulations}

In recent years, star cluster simulations have reached new levels of
power and sophistication. For example, we are now in a position to run
direct collisional $N$-body models of objects at the lower end of the
globular cluster mass function,  incorporating much realistic physics
(such as stellar and binary star evolution),
for example see Hurley et al. 2005, where the open cluster M67 was modelled.
\nocite{2005MNRAS.363..293H}  With this type of
modelling comes new challenges. One of the chief among these is the
question of what initial conditions should be used for direct,
realistic  globular cluster simulations. Of course, the initial
conditions adopted for any  given run depend strongly on the system
being modelled and the aims of the  simulation. Nevertheless, it is
important to develop a global understanding of how  initial conditions
affect subsequent evolution, in order that the {\it most suitable}
starting point can be selected for any given simulation: can we
constrain which initial conditions strongly affect  subsequent
long-term cluster evolution, and which are essentially irrelevant; or
observationally, do we see any objects which will evolve into globular
clusters  like those in the Galaxy over the next Hubble time? Examples
of modelling of globular clusters can be found in 
Phinney (1993)\nocite{1993ASPC...50..141P},
Druckier (1995)\nocite{1995ApJS..100..347D}, 
and Giersz \& Heggie (2003)\nocite{2003MNRAS.339..486G}. 

There are two ways to address this problem. From a modelling point of
view, as more and more large-scale simulations are calculated, with
varying initial conditions, it should become clear which parts of
parameter space (especially covering the IMF and initial spatial and
velocity structures such as mass segregation) strongly influence
cluster evolution, and how. Observationally, detailed measurements of
very young star clusters can help provide constraints on realistic
initial  conditions. Such measurements have already been utilized in a
number of studies -- for example, the direct modelling of LMC clusters
by \citep{mwd06} involved initial conditions strongly constrained by
the observed properties of young massive LMC objects such as R136,
NGC~1805, and NGC~1818.

\medskip

We propose the following MODEST questions:
\begin{itemize}
\item{In the next year: begin to run simulations designed specifically
to investigate the influence of initial conditions on the early,
intermediate, and late-time evolution of globular clusters. Can we
identify which are the most important initial conditions and which
have little or no effect? Can we infer initial conditions from
already-existing observations of young clusters?}
\item{In the next three to ten years: Can one run direct or
near-direct simulations with the aim of demonstrating whether the
super star clusters observed in starburst and interacting galaxies are
really globular cluster progenitors? If they are not, then what did
the Galactic globular clusters look like initially?}
\end{itemize}

\subsection{Are there black holes in globular clusters?}

Low--mass black holes (LMBHs) are expected as the  end products of the
evolution of stars populating the uppermost end of the IMF --  i.e.,
those with $M \geq 20 M_\odot$. If such remnants are formed without
significant initial velocity kicks, the retention fraction is expected
to be high  and the holes should
constitute a dynamically important cluster
sub-population  \citep{kul93,sig93}. Depending on the shape of the
upper IMF, most if not all globular clusters are expected to possess
this population of up to several hundred LMBHs early in their
life. Within $\sim 1$ Gyr of formation,  most LMBHs in a cluster have
settled via mass segregation to form a centrally  concentrated
core. Eventually this core is sufficiently dense that multiple-hole
interactions occur, resulting in the formation of BH-BH binaries and
the  ejection of single holes. Subsequent interactions harden the
BH-BH  binaries until eventually the recoil velocity is high enough
for ejection.  Several single holes are also expected to be ejected
during this  hardening process. Hence, it is thought that the LMBH
population in a  cluster completely depletes itself within a few
Gyr. Nonetheless, the  LMBH population is expected to inject
significant amounts of energy into  the stellar core in a cluster
before depletion, both through the  dynamical scattering of stars and
the removal of BH mass from the cluster centre. $N$-body simulations
show that this influence is in many cases enough to  significantly
alter (expand) the structure of the stellar core. Therefore, LMBHs
likely represent an important (and often neglected) dynamical
influence in the  early and intermediate phases of star cluster
evolution \citep{mwd06}.

Intermediate--mass black holes (IMBHs) in clusters are interesting
 because it is thought that such objects may represent the seeds of
 the super-massive black holes (SMBHs; $M > 10^6 M_\odot$)  which are
 inferred to exist both in high-redshift galaxies (where they are
 believed  responsible for quasars and AGN), and in the local universe
 at the centres of our Galaxy and M31. Stellar dynamical simulations suggest
 that IMBHs can be formed in very dense young globular clusters, via
 the process of {\it runaway merging}.  In such clusters, the
 core-collapse timescale for the most massive stars can be shorter
 than their main sequence lifetimes. Core collapse may initiate a
 rapid sequence of direct collisions between stars, leading to  the
 production of a merged object (possibly an IMBH) of mass $\sim 0.1\%$
 of the cluster mass \citep{spz02}.  It is likely the inspiral and
 destruction of a suitable cluster near a galactic centre (e.g., the
 Arches or Quintuplet in our Galaxy) may seed 
 or contribute to the growth of super-massive black holes 
($M > 10^6 M_\odot$) as described in Section 4.
%the bulge with an
% IMBH, which may subsequently accrete material and grow into  a
% SMBH. 
The possibility of detecting IMBHs in globular clusters  is
 thus both highly relevant and intriguing.

To date, we possess only indirect and/or debated evidence for IMBHs or
populations of LMBHs in globular clusters.  The presence of a $\sim
2500 M_\odot$  IMBH in the nearby globular cluster M15 \citep{ger02}
and a  $1.8\times10^4 M_\odot$ IMBH in the massive stellar cluster G1
in M31  \citep{geb05} have been inferred from HST
measurements. However, these detections are contested by
\citet{baum03a,baum03b}, whose $N$-body modelling  suggests that
neither detection {\it requires} the presence of an IMBH -- that is,
each set of measurements can seemingly be explained by models with
large central  populations of stellar remnants such as neutron stars
and white dwarfs. It has been suggested that detected X-ray emission
from G1 may be the result of accretion of gas by a central black hole
(Pooley \& Rappaport, 2006). \nocite{2006ApJ...644L..45P}  The more
general question of which globular clusters may contain IMBHs has also
been considered in Baumgardt et al (2005).
\nocite{2005ApJ...620..238B} The only  evidence for LMBH populations
in globular clusters is the observation  \citep{mg03a,mg03b} that
intermediate age clusters in the Magellanic  Clouds possess a wide
range of core sizes, and that this range apparently correlates  with
cluster age. This trend can, at least in part, be explained by the
dynamical  influence of LMBH populations \citep{mwd06}.

The main problem in detecting the presence of IMBHs (or centrally
concentrated LMBH populations) in globular clusters is that these
objects have only small spheres of dynamical influence: radius $\sim
0.1$ pc for a $2000 M_\odot$ IMBH in a typical cluster core. At
present therefore,  measurements are limited by resolution. This
problem will likely be solved on a time-scale of at least ten years,
with the advent of $50-100$m-class telescopes and functional adaptive
optics at visible wavelengths. Together these would permit full (3D)
dynamical studies of globular clusters out to the Magellanic Clouds
\citep{elt05}, allowing secure detections of IMBHs in globular cluster
cores, as well as the possibility of investigating the effects of LMBH
populations in intermediate-age clusters. In the meantime, ever more
realistic modelling can help place constraints on both IMBH and LMBH
formation in clusters,  as well as what observational signatures
should be expected from the presence of such objects.

\medskip

We therefore propose the following MODEST questions:
\begin{itemize}
\item{In the next year: using $N$-body modelling, can we make any new
predictions about the observational signature(s) of a population of
LMBHs in a globular cluster?}
\item{In the next three years: can we resolve the disagreement between
the observations of objects such as M15 and M31-G1 which infer the
presence of IMBHs, and $N$-body modelling which suggests IMBHs are not
required in order to explain the observed dynamics?  Can we refine
models of stellar evolution to make concrete predictions about whether
we should really be expecting IMBHs or large populations of LMBHs in
globular clusters?}
%\item{In the next ten years (or more): can we obtain irrefutable
%observational evidence  of an IMBH in at least one globular cluster?}
\end{itemize}

\section{Galactic nuclei}
\label{sec:galnuc}

\subsection{The final parsec problem}

It is generally accepted that hierarchical models best explain the
formation of structures in the Universe, down to the size of a galaxy
\citep{WR78,KWG93,SpringelEtAl05}. This means that in their lifetimes
galaxies typically merge with one or more other galaxies. A good
example of this is the Antenn{\ae} ``galaxy'', which actually consists
of {\em two} colliding galaxies, NGC~4038 and NGC~4039 \citep{WS95}.
Almost all galaxies appear to have a central supermassive black hole
(SMBH, \citealt{FF05} for a review), hence mergers of galaxies can
eventually lead to mergers of the SMBHs
\citep[e.g.,][]{MHN01,HK02,VHM03}.  At the beginning of the evolution,
dynamical friction makes the orbits of the SMBH decay, so that they
sink down to the centre of the merging system. 
 Strong interactions with surrounding stars
coming from the stellar system in which the SMBHs are embedded remove
energy and angular momentum from the SMBHs after they have formed a
bound binary system.  These stars are re-ejected into the stellar
system with an increased kinetic energy and thus the semi-major axis
of the SMBH binary shrinks.  The rate of shrinking slows down after
the SMBHs are close enough that they are more massive than the
enclosed stellar mass.  This typically happens at a separation $\sim
0.1-1$~pc, significantly larger than the $0.001-0.01$~pc needed so
that gravitational radiation alone can cause the binary to merge
within a Hubble time.  The ``final parsec problem"
\citep{BBR80,MM03,Yu02} thus consists of identifying processes that
can bring the binary separation from $\sim$1~pc to the realm of
significant gravitational radiation. The efficiency of such processes
has major implications for the growth and mergers of SMBHs, galaxy
evolution, and sources for future space-based gravitational wave
detectors such as the {\it Laser Interferometer Space Antenna} ({\it
LISA}). For a comprehensive review of the various aspects of formation
and evolution of binary MBHs we refer to \citet{MM05}.

In a galactic nucleus that has negligible amounts of gas and that has
an axially symmetric potential, individual stars in the nucleus
essentially conserve their angular momentum from orbit to orbit.
Thus, after the stars whose pericentres take them near the SMBH binary
are ejected from the system via a gravitational slingshot (hence the
``loss cone" is emptied; see \citealt{FR76}), further tightening of
the binary requires that distant two-body encounters between stars
send some of them on orbits radial enough to interact with the binary.
The timescale for this two-body relaxation can be billions of years,
but recent work based on the empirical relation between the mass of an
SMBH and the velocity dispersion or mass of its host galaxy bulge
\citep{FM00,Gebhardt00,MF01b,MF01,TremaineEtAl02} suggests that SMBH
binaries with total mass $M_{\rm BBH}<10^7\,M_\odot$ can be hardened
to merger in less than a Hubble time \citep{MM03}.  This is the realm
relevant to low-frequency gravitational wave detectors such as {\it
LISA}.  In contrast, it is still unclear whether a combination of
other factors such as gas drag \citep{EscalaEtAl05,MP05} or
triaxiality of the galactic nuclear potential
\citep{HolleyBockelManEtAl02,PM02,MP04} suffices to produce efficient
mergers for more massive SMBH binaries.

Ideally, one would like to ensure inclusion of all relevant physical
effects with a direct-summation N-body treatment of galactic centre
dynamics.  However, the actual number of stars in the central few
parsecs of a galaxy is $10^{7-8}$, which is too much to simulate in
this way even for special-purpose supercomputers such as the GRAPE-6,
which reach a computational power of 64
Tflops\footnote{\tt{http://grape.astron.s.u-tokyo.ac.jp/grape/}}.
Using fewer particles decreases the effective two-body relaxation time
and therefore introduces an artificially high rate of binary hardening.

A possible solution is parallel usage of a direct-summation N-body
code on a cluster of special-purpose GRAPE-6 nodes. However at the
present time, there is no N-body code which treats close encounters
rigourously (through Kustaanheimo-Stiefel two-body as well as chain
regularisation) and is both adapted for the use on GRAPE-6 and fully
parallelized \citep{Aarseth99,Aarseth03}.

Another possibility is that the real dynamics are not so sensitive to
the number of stars.  For example, it has been proposed that
non-axisymmetries could help the binary shrink much faster thanks to
the chaotic nature of the stellar orbits
\citep{HolleyBockelManEtAl02,PM02,MP04}.  Recent work based on this
approach suggests that the rate of orbital decay is roughly
independent of the total number of particles
\citep{HBS06,BerczikEtAl06}. However, caution is appropriate because
extrapolation to the much larger number of stars in real galaxies
requires theoretical scalings which are not fully understood. N-body
simulations should be therefore envisaged as a source of encouragement
and motivation rather than solid and robust proofs.

\medskip

For the MODEST questions on this subject we propose:

\begin{itemize}

\item In the next year: observationally, what is the dynamical state
of a galactic centre just after a major merger?  In particular, what
is the rotational structure, and how triaxial are the centres?

\item In the next three years: what is the influence of nuclear
rotation on the dynamics of a SMBH binary?  Initial investigations,
e.g., \citet{BerczikEtAl06}, suggest that the influence could be
substantial, but this needs to be coupled with observations.

\item In the next ten years: what is the evolution of the eccentricity
of an SMBH binary from its formation until it becomes detectable with
low-frequency gravitational radiation detectors such as {\it LISA}?
This will require incorporation of rotation, triaxiality, and the
effects of resonances on dynamical friction \citep{TW84}, and
substantial input from analytical treatments, direct N-body summation
techniques, and more approximate approaches.  The potential payoff is
that any residual eccentricity detected with {\it LISA} or similar
instruments might then be used to untangle important elements of the
mergers.

\end{itemize}

\subsection{Stellar dynamics around a massive black hole}
\label{subsec:Nucl_MBH}

The central SMBH and the stellar system interact through many channels
in  addition to the smooth gravitational potential. For example, stars
can produce gas to be accreted on to the SMBH, through normal stellar
evolution, collisions, or disruptions of stars by the strong central
tidal field. These processes may contribute significantly to the mass
of the SMBH. Tidal disruptions trigger phases of bright accretion that
may reveal the presence of an SMBH in an otherwise quiescent, possibly
very distant, galaxy. Collisions may create observationally peculiar
stellar populations. Also, stars too compact to be tidally disrupted
are swallowed whole if they are kicked directly through the horizon
(``direct plunges'') or progressively inspiral down to a relativistic
unstable orbit through emission of gravitational waves (GWs). The
latter process, known as an ``Extreme Mass Ratio Inspiral'' (EMRI)
will be one of the main targets of {\it LISA}.

Many different numerical schemes have been applied to the simulation
of galactic nuclei hosting an SMBH. Most of them rely on the
assumptions of an isolated, spherical system in dynamical equilibrium
(e.g., direct integration of Fokker-Planck equation by
\citealt{MCD91}, Monte-Carlo methods by \citealt{DS82} and
\citealt{FB02b}, and gas-dynamical treatment by \citealt{ASFS04}). In
these cases, only ``collisional'' effects can bring stars on to the
loss-cone, i.e., the very elongated orbits which allow close
interaction between a star and the SMBH. These effects include
(diffusive) 2-body relaxation, large-angle scatterings, direct
collisions, and resonant relaxation. The approximate methods just
mentioned generally include only diffusive relaxation and, in rare
cases, collisions and large-angle scatterings. There have only recently
been direct N-body simulations of clusters with a central object
\citep[][in particular]{BME04a,BME04b,PMS04}.

As with binary SMBH simulations, for these applications N-body
simulations are invaluable because they dispense with the necessity of
most approximations but it seems unlikely that they will completely
supersede other approaches in the next few years. Their most obvious
limitation is the steep computing time scaling $t_{\rm CPU}\propto
N_{\rm part}^{2-3}$ which currently limits the number of particles to
about $10^6$, short of the $10^7-10^8$ stars in even small galactic
nuclei. More fundamentally, the community is missing an algorithm
suitable to tackle this particular class of problems. The integration
of millions of nearly Keplerian orbits with the Hermite scheme used in
usual N-body codes causes spurious change in the orbital constants and
may, for instance, lead to incorrect star-MBH interaction rates due to
eccentricity increase. Therefore, numerical errors dominate the
effects of the actual star-star perturbations.

The development of an N-body regularisation scheme to integrate the
motion and mutual perturbations of a large number of light objects
orbiting the same massive body has become a challenging priority. The
chain scheme is not suitable here because it does not allow
regularisation of the interaction of the SMBH with each close star
simultaneously (see \citealt{Aarseth03b} for technical background).
Some inspiration may be provided by the symplectic codes in use in
planetary dynamics \citep{WH91,WHT96}.  These codes are  optimised for
nearly-Keplerian orbits with a single dominant object, and bound the
energy error over large numbers of orbits.  They may therefore have
applicability to the central regions of galactic nuclei that are
dominated by an SMBH.

It is, however, likely that new development will be necessary for
dynamics around an SMBH because excellent accuracy in all the orbital
elements, not just the semimajor axis, may well be necessary to
preserve the proper interactions with narrow resonances or to include
the correct effects of processes such as general relativistic
precession.

The inclusion of relativistic contributions to dynamics is  especially
important for the proper treatment of effects such as resonant
relaxation \citep{RT96} or Kozai cycles
\citep{Kozai62,LZ76,IZMV97,MH02b}, which depend on the persistence of
certain phase relations over hundreds of orbits or more.  Even
relatively small effects can have an influence over this many orbits,
hence precision of integration and inclusion of relativity are at a
premium.  In addition, estimates of the distributions of mass ratio,
eccentricity, etc., are needed to construct reliable template banks
for detection of the gravitational radiation from EMRIs. Simulations
based on purely Newtonian schemes may well lead to completely
incorrect results for these purposes. General relativistic effects up
to the 2.5 post-Newtonian order (and hence including both precession
and radiation reaction) have been implemented in the codes HNBody
\citep{GMH06} and N-body6++ \citep{KupiEtAl06}, but the applications
are in their infancy.

Whatever the eventual solution, it is likely that such N-body codes
will initially lack the efficiency to treat more than $10^4-10^5$
stars, corresponding to the inner part of the influence region of a
realistic nucleus. An important first step would be to consider the
central SMBH as a fixed particle, i.e., treat it as an external
potential. Neglecting the random motion of the SMBH may have
consequences for, e.g., the rate of tidal disruptions because this
motion should allow loss-cone replenishment. However, this should only
become an issue when a region larger than the influence radius can be
simulated. Furthermore, this idealisation will prove useful for
comparison with analytical and approximate numerical approaches which
generally rely on it. It is also the best way to assess later the role
of the SMBH motion by comparison with more realistic simulations.

To pave the way for N-body studies and to complement them, more
approximate but much faster and more flexible methods are
invaluable. Monte Carlo (MC) statistical approaches seem ideal
because, being based on particles, they make it easier to follow the
evolution of individual orbits and to include individual star-star or
star-SMBH interactions. A promising avenue would be to combine aspects
of the MC approach of \citet{HA05,HA06} to that first pioneered by
\citet{Henon73} for globular cluster dynamics. \citeauthor{HA05}
followed the orbital evolution of test particles due to diffusive and
resonant relaxation and to GW emission in the Keplerian potential of
the SMBH assuming a fixed stellar background. The H\'enon approach
would evolve the stellar distribution self-consistently --thus
obtaining the correct mass-segregation effects-- but its only
application to galactic nuclei so far \citep{FB02b} lacks the ability
to resolve the dynamics of stars in or near the loss cone on
satisfactorily short timescales. Shapiro and collaborators
\citep{DS82,Shapiro85} developed an MC code which represents the
cluster as a set of spherical shells like in the H\'enon scheme but
where the effects of relaxation is computed by explicit integration of
diffusion coefficients like in direct Fokker-Planck codes, rather than
pairwise interactions between particles. This method had the important
advantages of allowing one to increase the resolution in the central
regions and to follow particles very close to the SMBH on orbital
timescales, but it has only been used for single-mass cases. Such an
algorithm, if it can be extended to a stellar mass spectrum, would be
an extremely useful tool.

Theoretical predictions for rates and characteristics of EMRI events
have proven quite problematic so far (see \citealt{Sigurdsson03} for a
quick review and \citealt{HA05,HA06} for new developments). Even in
the standard case where only diffusive relaxation on single stars is
considered, different authors find results scattered over such a large
range that it is not clear whether only a few events will be detected
with {\it LISA} \citep{HA05} or whether they will turn out to be an
embarrassment of riches, preventing the individual detection of each
other and of other sources \citep{BC04}. Recently, non-standard
processes to bring compact objects very close to an SMBH have been
suggested. These include the accumulation  of red-giant cores by tidal
peeling \citep{DK05}, tidal separation of binaries \citep{MFHL05},
stellar formation in an accretion disk \citep{Levin03,Nayakshin05} and
dissipative interactions with a disk \citep{SKH04}. In all these
cases, the orbital evolution will start being dominated by GW emission
(as opposed to relaxation) at a smaller semimajor axis and much
smaller eccentricity than in the ``standard'' case and the EMRI should
be very nearly circular in the {\it LISA} band. The study of these
various types of EMRIs will be one of the main applications of the
numerical methods envisioned here.

Noticeably each of these processes was (also or uniquely) offered as a
way to explain the origin of the young massive ``S'' stars orbiting
{\SgrA} \citep[e.g.,][]{GenzelEtAl03}. Hence it is no coincidence
that, unless they somehow get natal kicks comparable to their orbital
speeds of thousands of kilometers per second, the remnants of the
``S'' stars will have the appropriate orbital parameters to become
EMRIs, a fact with strong bearing on {\it LISA} detection rates if the
situation around {\SgrA} is typical.

It has also been suggested that intermediate-mass black holes (IMBH,
with masses $\sim 10^{2-4}\,M_\odot$) formed in young massive star
clusters within $\sim 100$~pc of the centre of a galaxy could sink to
the centre and merge with an SMBH  \citep{Miller05,PZetAL06,MME06}.
If this happens it would be an extremly strong source for {\it LISA},
with unique potential for mapping the spacetime around a rotating SMBH
\citep{Miller05}.  Currently, however, there are many uncertainties
about the various steps in the sinking process, from the settling of a
cluster in a galactic nucleus to the stripping of that cluster to the
processes that allow an IMBH to merge with an SMBH (for example, is
there any stalling and if so, will other IMBHs come in and cause
mutual ejection?).

Finally, while structures such as triaxial bulges, bars or stellar
discs are common on scales of 100--1000\,pc, the influence of
non-sphericity at small and intermediate scales on the structure and
evolution of the nucleus has been little explored. The existence of a
large fraction of ``centrophilic'' (box and chaotic) orbits in triaxial
structures has the potential of boosting the rate of star-SMBH
interactions by orders of magnitude
\citep{HolleyBockelManEtAl02,PM02,MP04,HBS06}. For EMRIs, though, it
is not clear whether such orbits, with very large initial semimajor
axis and eccentricity, have a chance to shrink to {\it
LISA}-detectable frequencies without being perturbed into a direct
plunge or a wider orbit.

\medskip

We therefore propose the following MODEST questions:

\begin{itemize}

\item In the next year: are the basic codes used to calculate EMRI
rates consistent with each other? As proposed to us by Richard
Mushotzky, we suggest that a precisely defined test case, accessible
to direct N-body summation methods as well as to statistical
approaches, should be simulated by several independent groups. Such
comparison has proved very enlightening in the case of the
``collaborative experiment'' in cluster dynamics organised by Douglas
Heggie (\citealt{Heggie03}, see also
{\tt{http://www.manybody.org/modest/WG/wg7.html}}) or in the field of
cosmological hydrodynamics simulations \citep{FrenkEtAL99}.  For
example, current codes could treat a cluster of $10^4$ point-mass
single stars hosting a central SMBH with a mass 1\% of the
total. Several quantities could be compared, including the time
evolution of the capture rate. The observed similarities and
differences could help guide further treatments. For the number of
EMRIs to be significant with such a particle number, the cluster needs
to be made more compact than any known real system, to boost
relativistic effects relative to relaxation. Although lacking physical
realism, this setting will allow one to test and calibrate the approximate
methods by comparison with direct N-body.

\item In the next three years: what are the capture and tidal
destruction rates implied by the actual distribution of stars around
the SMBH in our own Galaxy?  The radial dependence of number density
is reasonably well constrained within $\sim$0.01~pc of the centre, a
region containing some $10^{4-5}$ stars, so a specific simulation over
the  $\sim 10^9$~yr relaxation time would be informative.

\item In the next ten years: what is the true influence of
nonaxisymmetry at large distances on the inner few parsecs, where
EMRIs interact?  Does this lead to large rates of {\it LISA}
detections, or does it instead produce direct plunges?  Before direct
N-body methods are able to deal with $>10^7$ particles, the
relaxational dynamics of non-spherical systems could be studied with
hybrid schemes borrowing from ``collisionless'' N-body and
Fokker-Planck or MC codes, an option still virtually unexplored (with
the exception of \citealt{JSH99}).  As a separate but related matter,
what are the processes that lead to an IMBH-SMBH merger, and what is
the expected rate of {\it LISA} detections?

\end{itemize}

\subsection{Stellar collisions}

Galactic nuclei are one of the few environments in which collisions
involving single stars should occur on a relatively short timescale.
For instance, within $\sim 0.03\,$pc of {\SgrA}, a $1\,\Msun$ main
sequence (MS) star should experience, on average, one collision in
less than 10\,Gyr. For a giant, this timescale is reduced to a few
$10^7$\,yr. Although collisions probably do not strongly influence the
stellar dynamics, they are of great interest as a way to produce
unusual stellar populations. They have been suggested as the cause of
the apparent paucity of giants in the vicinity of {\SgrA}, although
giants irradiated by the X-ray radiation of {\SgrA} may actually
masquerade as massive MS stars \citep{JimenezEtAl06}.

Collisions in galactic nuclei occur at relative speeds of a few
100\,{\kms} or higher, making mergers an unlikely outcome for low-mass
stars.  In particular the possibility of growing ``super blue
stragglers'' through a sequence of collisions seems excluded. The mass
and energy loss for such high-velocity collisions between MS stars has
been studied exhaustively \citep{FB05} but much remains to be done for
the more likely case of a collision between a giant and a more compact
object.  In relatively small galactic nuclei (typically hosting an
SMBH less massive than $10^7\,\Msun$), collisions involving stellar
BHs are also of special importance because mass segregation probably
concentrates these objects around the central SMBH \citep{FASK06}.

High-velocity collisions between a giant and a smaller star were
computed, using SPH, by \citet{BD99} who found that during a typical
collision the impactor, flying through the giant's envelope, causes
only relatively little mass loss; the giant is likely to recover on a
short (thermal) timescale. Complete removal of the envelope can only
happen if the smaller star is captured and a common-envelope (CE)
system is formed, an outcome too rare to explain the dearth of giants
at the Galactic centre. On the other hand, \citet{DBBS98} showed that
collisions between giants and binary stars may be more efficient at
depleting the giant population, either by creating CE systems or by
ejecting the giant's core from its envelope, if binaries are common
enough. However, except for the detection of transient X-ray sources
at $< 1\,$pc from {\SgrA} \citep{MunoEtAl05}, very little is known
about binary populations in galactic nuclei.

\medskip

Our MODEST questions are:

\begin{itemize}

\item In the next year: what are the dynamics of binaries in galactic
nuclei?  The central goal is to study the survival of binaries in an
environment with a much higher velocity dispersion than exists in
globular clusters.  It will be particularly important to study this
question using stellar dynamical simulations with a large number of
particles (see Sec.~\ref{subsec:Nucl_MBH}).

\item In the next three years: what is the evolution and appearance of
a giant star whose envelope has been partially removed by a collision
(or a strong tidal interaction with the SMBH, see
\citealt{DSGMG01,DK05}). Also, what is the evolution of common
envelope binaries formed through red giant collisions?   These systems
may be the progenitors of compact binaries, possibly explaining (some
of) the X-ray sources observed around {\SgrA}.

\item In the next ten years: what are the hydrodynamic and possibly
magnetohydrodynamic results of collisions between giants and smaller
stars, and between compact objects (especially stellar-mass black
holes) and extended stars?  For the former, simulations need to cover
a much more extended region of parameter space (masses, evolutionary
stage of the giant, relative velocity and impact parameter) than
published so far.  For the latter, it will be important to understand
how damaging the collisions are, and how much mass the compact star
can accrete.

\end{itemize}

\bigskip
%\ackowledgments

\noindent {\bf {\Large Acknowledgments}}

\noindent
The authors acknowledge the input provided by other members of the
MODEST collaboration, in particular those who attended the Modest-6A
workshop in Lund.  JED acknowledges support from the UK Particle
Physics and Astronomy Research Council via the University of
Leicester's theoretical astrophysics rolling grant.  MBD is a Royal
Swedish Academy Research Fellow suuported by a grant from the Knut and
Alice Wallenberg Foundation.  ADM gratefully
acknowledges support in the form of a PPARC Postdoctoral Fellowship.
MCM was supported in part by the
Research Associateship Programs Office of the National Research
Council and Oak Ridge Associated Universities. 
The work of PAS has been supported in the framework of the Third
Level Agreement between the DFG (Deutsche Forschungsgemeinschaft) 
and the IAC (Instituto de Astrof\'\i sica de Canarias).
SPZ was supported by
the Royal Netherlands Academy of Arts and Sciences (KNAW), by the
Leids Kerkhoven Bosschafonds (LKBF) by the Netherlands Research School
for Astronomy (NOVA) and by the Netherlands Organization for
Scientific Research (NWO, under grant \#635.000.001 and \#643.200.503).

\bigskip

\end{document}